\documentclass[aps,prc,preprint,superscriptaddress,showpacs,preprintnumbers]{revtex4}

\usepackage{graphicx}
\usepackage{dcolumn}

\begin{document}

\title{Nature of yrast excitations near $N=40$: Level structure of $^{67}$Ni}
\author{S. Zhu}
\author{R. V. F. Janssens}
\author{M. P. Carpenter}
\affiliation{Physics Division, Argonne National Laboratory, Argonne,
Illinois 60439, USA}
\author{C. J. Chiara}
\affiliation{Physics Division, Argonne National Laboratory, Argonne,
Illinois 60439, USA} \affiliation{Department of Chemistry and
Biochemistry, University of Maryland, College Park, Maryland 20742,
USA}
\author{R. Broda}
\affiliation{Institute of Nuclear Physics, Polish Academy of
Sciences, PL-31342 Krakow, Poland}
\author{B. Fornal}
\affiliation{Institute of Nuclear Physics, Polish Academy of
Sciences, PL-31342 Krakow, Poland}
\author{N. Hoteling}
\affiliation{Physics Division, Argonne National Laboratory, Argonne,
Illinois 60439, USA} \affiliation{Department of Chemistry and
Biochemistry, University of Maryland, College Park, Maryland 20742,
USA}
\author{W. Kr{\'o}las}
\affiliation{Institute of Nuclear Physics, Polish Academy of
Sciences, PL-31342 Krakow, Poland}
\author{T. Lauritsen}
\affiliation{Physics Division, Argonne National Laboratory, Argonne,
Illinois 60439, USA}
\author{T. Paw{\l}at}
\affiliation{Institute of Nuclear Physics, Polish Academy of
Sciences, PL-31342 Krakow, Poland}
\author{D. Seweryniak}
\affiliation{Physics Division, Argonne National Laboratory, Argonne,
Illinois 60439, USA}
\author{I. Stefanescu}
\affiliation{Physics Division, Argonne National Laboratory, Argonne,
Illinois 60439, USA} \affiliation{Department of Chemistry and
Biochemistry, University of Maryland, College Park, Maryland 20742,
USA}
\author{J. R. Stone}
\affiliation{Department of Chemistry and Biochemistry, University of
Maryland, College Park, Maryland 20742, USA} \affiliation{Department
of Physics, University of Oxford, OX1 3PU Oxford, UK}
\author{W. B. Walters}
\affiliation{Department of Chemistry and Biochemistry,
University of Maryland, College Park, Maryland 20742, USA}
\author{X. Wang}
\altaffiliation[Present address: ]{Department of Physics, Florida
State University, Tallahassee, Florida 32306, USA}
\affiliation{Physics Division, Argonne National Laboratory, Argonne,
Illinois 60439, USA} \affiliation{Physics Department, University of
Notre Dame, Notre Dame, Indiana 46556, USA}
\author{J. Wrzesi{\'n}ski}
\affiliation{Institute of Nuclear Physics, Polish Academy of
Sciences, PL-31342 Krakow, Poland}

\date{\today}

\begin{abstract}
Excited states in $^{67}$Ni were populated in deep-inelastic
reactions of a $^{64}$Ni beam at 430 MeV on a thick $^{238}$U
target. A level scheme built on the previously known 13-$\mu$s
isomer has been delineated up to an excitation energy of ~5.3 MeV
and a tentative spin and parity of (21/2$^-$). Shell model
calculations have been carried out using two effective interactions
in the $f_{5/2}pg_{9/2}$ model space with a $^{56}$Ni core.
Satisfactory agreement between experiment and theory is achieved for
the measured transition energies and branching ratios. The
calculations indicate that the yrast states are associated with
rather complex configurations, herewith demonstrating the relative
weakness of the $N=40$ subshell gap and the importance of multi
particle-hole excitations involving the $g_{9/2}$ neutron orbital.
\end{abstract}


\maketitle
\section{INTRODUCTION}
The presence of shell gaps with magic numbers of nucleons is a
cornerstone of nuclear structure. Over the past decade it has
increasingly become clear that magic numbers are not immutable, but
depend on the ratio of protons and neutrons \cite{Sor08,Jan09}. In
discussions of magic numbers, neutron number $N=40$ has historically
been a subject of debate, especially in the case of the Ni isotopes.
Proton number $Z=28$ is magic, and, for neutrons, a sizable energy
gap at $N=40$ is thought to separate the $pf$ shell from the
intruder $g_{9/2}$ state, potentially making $Z=28$, $N=40$
$^{68}$Ni a doubly-magic nucleus. Experimentally, the occurrence of
shell closures in $^{68}$Ni was first suggested based on the
observation of a 1770-keV 0$_2^+$ level as the lowest excited state,
followed by a 2$_1^+$ state of relatively high excitation energy
(2034 keV) \cite{Ber82,Bro95}. The discovery of several isomeric
states in $^{68}$Ni and in neighboring nuclei \cite{Grz98, Ish00}
supported the case for its magic character further, as did the
results from Coulomb excitation measurements indicating a $B(E2,
2_1^+\rightarrow0^+)$ reduced transition probability roughly three
times smaller than the corresponding value for $^{56}_{28}$Ni$^{
}_{28}$ \cite{Sor02,Bre08}. However, based on recent high-precision
mass measurements in the neutron-rich Ni isotopes (up to $^{73}$Ni),
the $N=40$ shell closure appears to be more doubtful when inferred
from changes in the two-neutron separation energies
\cite{Gue07,Rah07}. It has been argued in the literature that the
apparent contradiction between the $B(E2)$ value and the separation
energy is a consequence of the parity change across the $N=40$ gap,
with a sizable fraction of the low-lying $B(E2)$ strength residing
in excited states around 4 MeV, and the 2$_1^+$ level being
associated predominantly with a neutron-pair excitation
\cite{Sor02,Lan03}. The size of the $N=40$ gap is then of the order
of 2 MeV only and the corresponding discontinuity in the sequence of
orbitals corresponds at most to a subshell closure.

Beta-decay studies \cite{Han99,Mue99,Sor03,Gau05} and in-beam
investigations at intermediate beam energies \cite{Ald08, Gad10,
Lju10, Rot11} provide evidence for the onset of collectivity and
strong polarization of the $^{68}$Ni core in neighboring nuclei of
the region. For example, the 2$_1^+$ levels in $N=40$ $^{64}$Cr
\cite{Gad10} and $^{66}$Fe \cite{Han99} are located at excitation
energies as low as 420 and 573 keV, respectively. More generally,
the available low-spin level structures in these nuclei suggest
sizable admixture of spherical and deformed components in the
configurations near their ground states. These observations,
combined with the large $B(E2, 2_1^+\rightarrow0^+)$ value measured
for $^{70}$Ni$_{42}$ \cite{Per06}, lead to the conclusion that any
``island" of nuclei with indications of a significant $N=40$ gap is
rather localized. A main contributor to this situation is the
monopole tensor force \cite{Ots01,Ots05} between protons in the $pf$
shell and $g_{9/2}$ neutrons, where the occupation of the latter
orbital leads to the onset of deformation, as evidenced, for
example, by the presence of rotational bands in neutron-rich
$^{55-57}$Cr and $^{59-61}$Fe nuclei
\cite{Dea05,Zhu06,Dea11,Dea07,Hot08}. On the other hand, couplings
of protons and/or neutrons to the $^{68}$Ni core do not always
result in a large polarization of the core. For example, the first
excited state above the (19/2$^-$) isomer in $^{71}$Cu is located
2020 keV higher in energy, herewith mirroring the location of the
2$_1^+$ level in $^{68}$Ni \cite{Ste09}.

From the considerations above, it is clear that a satisfactory
description of nuclear structure in this mass region is still
lacking. This is also reflected in on-going theoretical efforts to
determine the most appropriate interactions for use in the
calculations. In this context, the present data on the neutron-hole
nucleus $^{67}$Ni provide an opportunity to test the most modern
interactions while investigating the nature of yrast excitations up
to moderate spin.

At present, only limited information is available on the low-lying
structure of $^{67}$Ni. Using deep-inelastic reactions of a
$^{64}$Ni beam at 350 MeV on a thick $^{208}$Pb target, Paw{\l}at
{\it et al.} \cite{Paw94} identified a 1008-keV isomeric state with
a half-life of T$_{1/2} > 0.3$ $\mu$s decaying through coincident
314- and 694-keV transitions towards the $^{67}$Ni ground state. The
presence of the isomer was later confirmed in a fragmentation
measurement where a 13.3(2)-$\mu$s half-life was determined
\cite{Grz98}. Feeding of the isomer in $^{67}$Co $\beta$ decay was
subsequently reported \cite{Wei99}. These three studies
\cite{Grz98,Paw94,Wei99} proposed spin and parity quantum numbers of
9/2$^+$ for the long-lived state and associated this level with the
occupation of the $g_{9/2}$ orbital by a single neutron. With the
NMR/ON technique, the magnetic dipole moment of the 1/2$^-$ ground
state was measured to be +0.601(5) $\mu_N$, a value differing only
slightly from the $\nu p_{1/2}$ single-particle value; a fact
regarded as evidence for the strength of the $N=40$ shell closure
\cite{Rik00}. For the isomer, a quenched $g$-factor value of
$|g|$=0.125(6) was reported in Ref.~\cite{Geo02} and was interpreted
as evidence for a 2\% admixture of a $\pi (f^{-1}_{7/2}f^{
}_{5/2})_{1+}\nu g_{9/2}$ configuration involving a proton
excitation across the $Z=28$ gap into the supposedly pure $\nu
g_{9/2}$ state.

Prior to the present work, no transitions feeding the isomeric state
had been reported. Here, despite the long half-life, eight new
states have been placed above the isomer from an investigation of
prompt-delayed coincidence events in a deep-inelastic reaction with
a pulsed beam.

\section{EXPERIMENT}
A number of experiments have demonstrated that the yrast states of
hard-to-reach neutron-rich nuclei can be populated in deep-inelastic
processes at beam energies $15\%-25\%$ above the Coulomb barrier
\cite{Jan02,For05,Bro06}, allowing experimental access to high-spin
structures in regions inaccessible with conventional heavy-ion
induced, fusion-evaporation reactions.

The experiment was carried out with a $^{64}$Ni beam delivered by
the ATLAS superconducting linear accelerator at Argonne National
Laboratory. The 430-MeV beam energy was chosen to correspond roughly
to an energy of 20$\%$ above the Coulomb barrier in the middle of a
55-mg/cm$^2$ thick $^{238}$U target. The beam was pulsed with a
412-ns repetition rate, each beam pulse being $\sim$$0.3$ ns wide.
Gammasphere \cite{Lee90}, with 100 Compton-suppressed HPGe
detectors, was used to collect events with three or more $\gamma$
rays in coincidence. The data were sorted into two-dimensional
($E_\gamma-E_\gamma$ matrices) and three-dimensional
($E_\gamma-E_\gamma-E_\gamma$ cubes) histograms under various timing
conditions. The prompt $\gamma\gamma\gamma$ cube (PPP cube) was
incremented for $\gamma$ rays observed within $\pm$20 ns of the beam
burst while, in the delayed $\gamma\gamma\gamma$ cube (DDD cube),
the transitions were required to occur in an interval of $\sim$40 to
$\sim$800 ns after the prompt time peak (excluding the subsequent
beam pulse), but within $\pm$20 ns of each other. In this way,
events associated with isomeric deexcitations could be isolated and
identified. The prompt-delayed-delayed (PDD) and
prompt-prompt-delayed (PPD) cubes were incremented by combining
prompt and delayed events. These proved critical in identifying
prompt $\gamma$ rays feeding isomeric levels as they revealed
themselves in double-gated spectra on the known transitions below
the isomer in the PDD cubes. The relations between these prompt
$\gamma$ rays were subsequently established by examining proper
double coincidence gates in the PPD and PPP cubes. Examples and
further details of this technique can be found in Refs.
\cite{Hot08,Ste09}.

The spins and parities of the levels were deduced from an
angular-correlation analysis. In addition, considerations based on
the fact that the reactions feed yrast states preferentially, and/or
on comparisons with shell-model calculations, were also taken into
account. The projectile-like products of deep-inelastic reactions
are usually characterized by no, or very little, alignment.
Therefore, the analysis of $\gamma$$\gamma$ angular correlations for
selected pairs of transitions is required \cite{For05,Hot06}. In
practice, in order to avoid as much as possible ambiguities in the
spin assignments, at least one known stretched transition was
included in the analysis.

\section{RESULTS}
\begin{figure}
        \centering
        \includegraphics[angle=-90,width=0.95\textwidth]{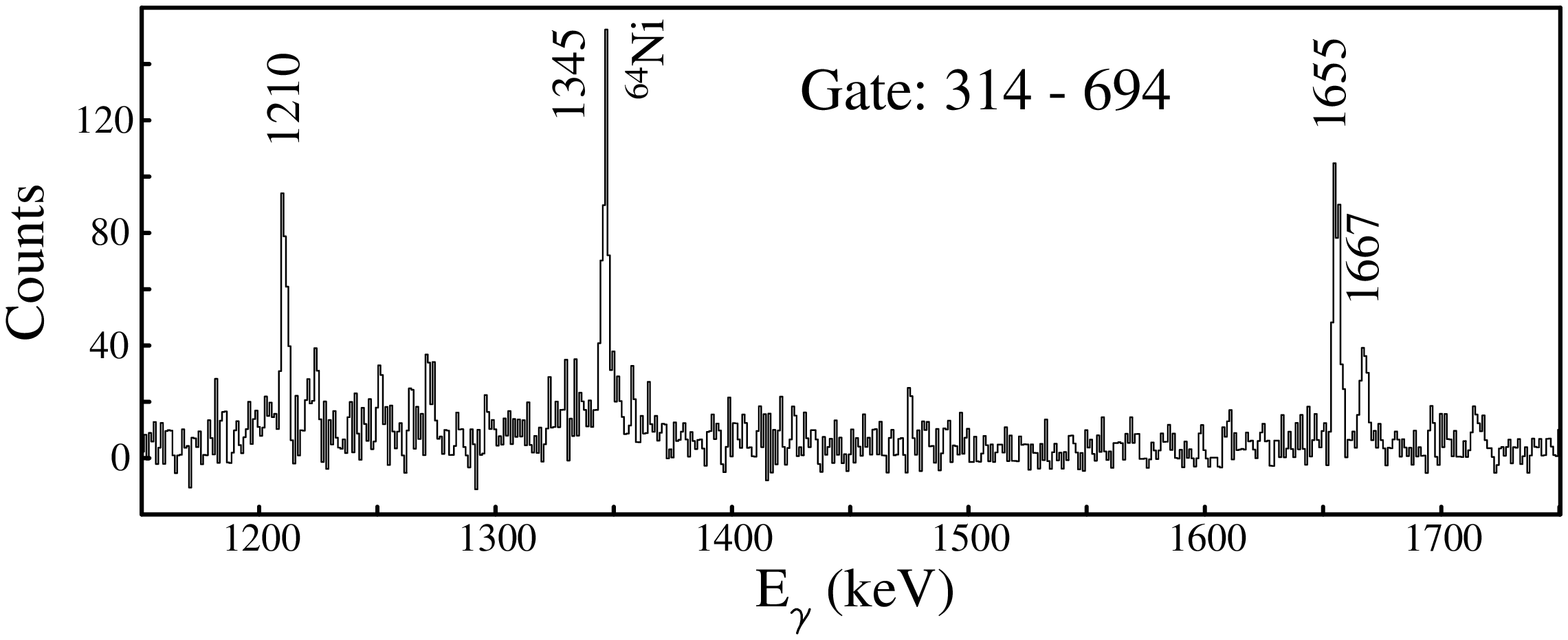}
        \caption{Partial spectrum with double coincidence gates set
        on the 314- and 694-keV cascade below the $^{67}$Ni 9/2$^+$
        isomer in the PDD cube. The spectrum shows three strong
        transitions feeding the 9/2$^+$ state; the 1345-keV $\gamma$
        ray of $^{64}$Ni is due to random coincidences (see text for
        detail).}
        \label{fig:fig01}
    \end{figure}

\begin{figure}
        \centering
        \includegraphics[width=0.8\textwidth]{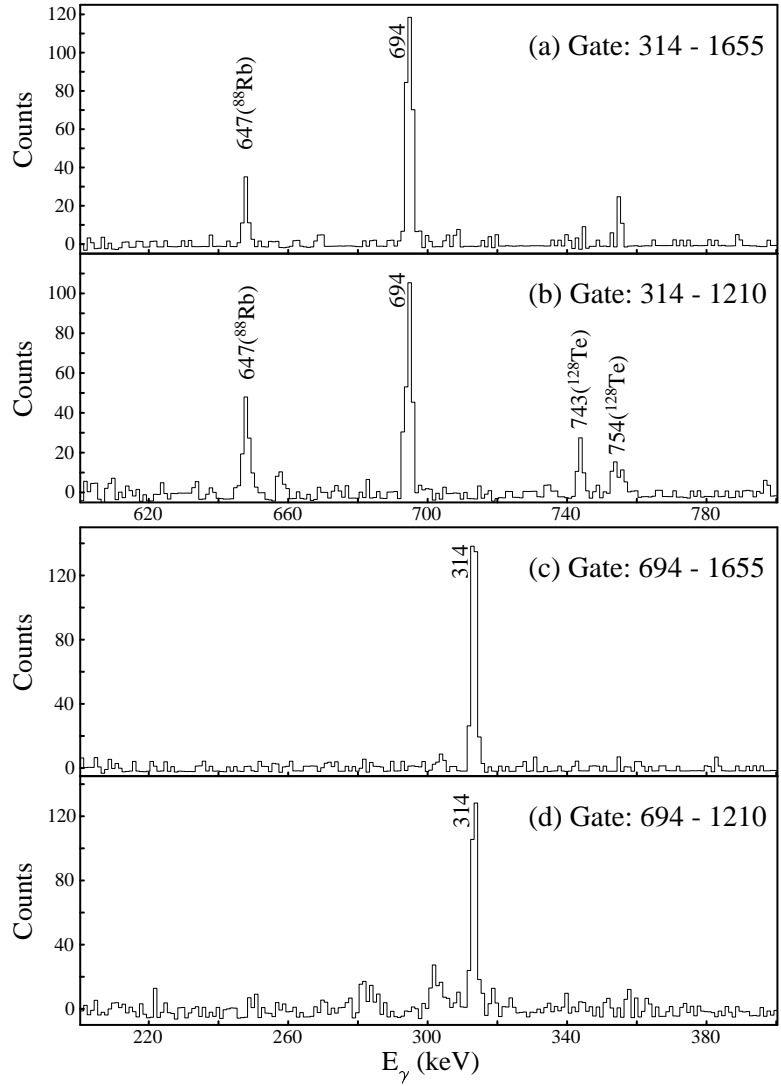}
        \caption{Partial coincidence spectra with different gates in the PDD
        cube establishing the feeding of the 9/2$^+$ isomeric state,
        see text for details. Note the change in energy scales
        between panels (a, b) and (c, d).}
        \label{fig:fig02}
    \end{figure}

In previous studies \cite{Paw94,Grz98,Wei99}, the 314- and 694-keV
transitions deexciting the 13.3(2)-$\mu$s isomer in $^{67}$Ni were
assigned to the $9/2^+$$\rightarrow$$5/2^-$$\rightarrow$$1/2^-$
cascade and no transition above this long-lived state was reported.
As stated above, transitions feeding the isomer were initially
identified in the present work by using the PDD coincidence data. A
coincidence spectrum from this PDD cube with double gates placed on
the 314- and 694-keV transitions is presented in
Fig.~\ref{fig:fig01}. Besides a 1345-keV line belonging to the
$2^+\rightarrow 0^+$ transition in $^{64}$Ni, three $\gamma$ rays
are clearly visible at 1210, 1655, and 1667 keV. The 1345-keV line
originates from Coulomb excitation of the $^{64}$Ni beam, and is
attributed to random coincidences. By double gating the PDD cube
with one of the newly discovered prompt $\gamma$ rays and one of the
delayed transitions, their mutual coincidence relationships can be
verified further. The results, displayed in Fig.~\ref{fig:fig02},
establish the feeding of the $^{67}$Ni isomer by the 1210- and
1655-keV transitions. Finally, additional evidence was provided by
the analysis of the PPD cube, where a double gate on the prompt
1655- and 1667-keV $\gamma$ rays yields a delayed spectrum in which
the 314- and 694-keV lines appear, consistent with the expected
coincidence relationships for $\gamma$ rays across the isomer. This
observation also implies that the 1655- and 1667-keV transitions are
in mutual, prompt coincidence.

\begin{figure}
        \centering
        \includegraphics[angle=-90,width=0.8\textwidth]{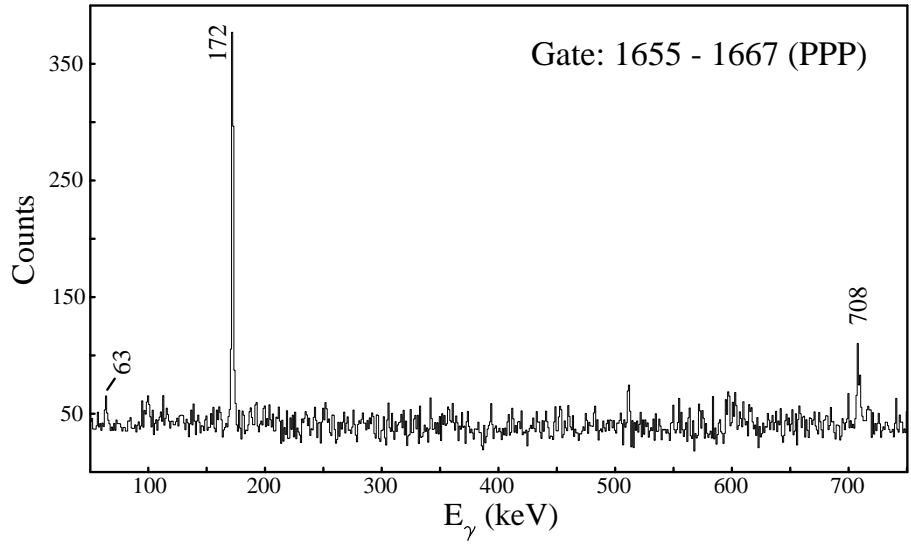}
        \caption{Partial coincidence spectrum with double gates set on the
        1655- and 1667-keV lines in the PPP cube showing the transitions
        from the states with highest excitation energy observed in the
        present work.}
        \label{fig:fig03}
    \end{figure}

\begin{figure}
        \centering
        \includegraphics[height=0.8\textheight]{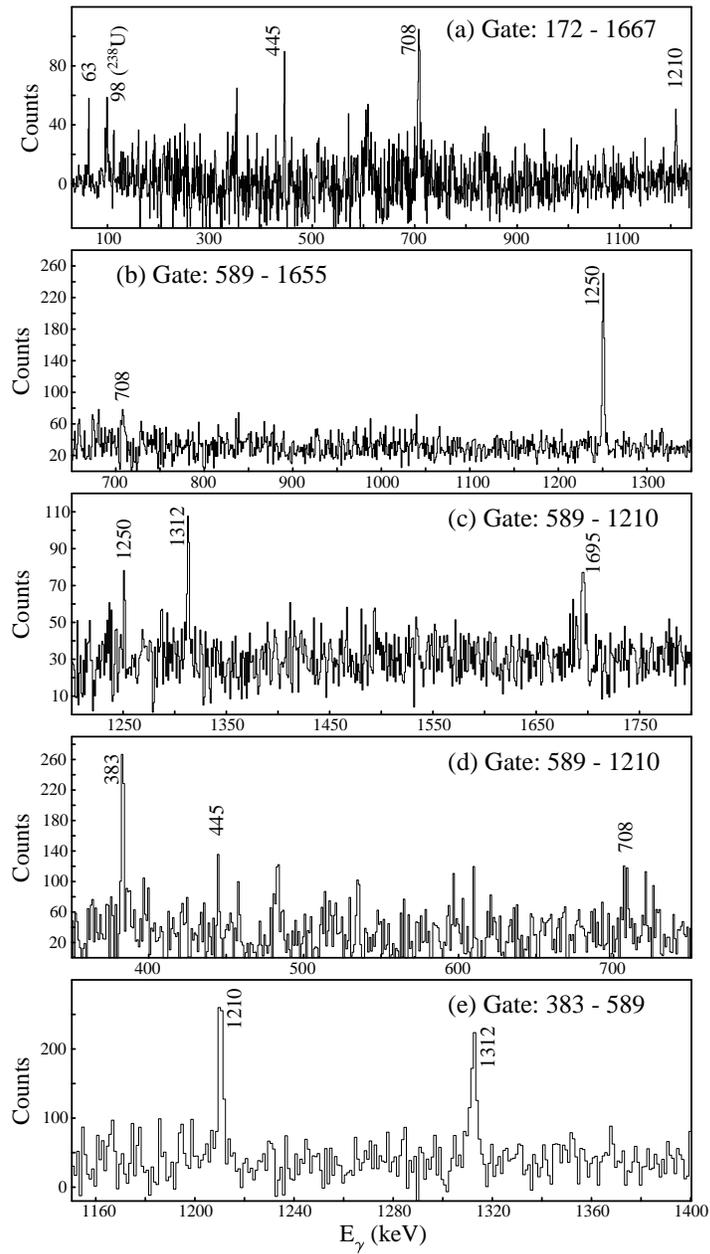}
        \caption{Partial spectra with different gates in
        the PPP cube demonstrating various coincidence relationships
        used to establish the level scheme of Fig.~\ref{fig:fig05}.}
        \label{fig:fig04}
    \end{figure}

Levels above the $^{67}$Ni isomer were investigated further in the
PPP cube with the newly-observed 1210-, 1655- and 1667-keV
transitions as a starting point. A double gate on the latter two
$\gamma$ rays reveals the presence of three additional lines at 63,
172, and 708 keV (see Fig.~\ref{fig:fig03}). The 1655- and 1210-keV
transitions are not in mutual coincidence, herewith establishing the
presence of parallel decay sequences. Exploiting additional
coincidence relationships, such as those displayed in
Fig.~\ref{fig:fig04}, it was possible to propose the level scheme of
Fig.~\ref{fig:fig05}. Thus, states at 2218, 2663, 3530, 3913, 4330,
and 4502 keV were firmly established through the various competing
decay paths. The ordering of the highest levels at 4565 and 5273 keV
is based on the measured $\gamma$-ray intensities. The presence of a
low-energy transition of 63 keV might suggest a longer half-life for
the 4565-keV state. Unfortunately, at this energy, the timing signal
of the large-volume germanium detectors is rather poor. This fact,
combined with the rather small intensity, made it impossible to
obtain firm information on the level lifetime. However, time spectra
gated on the transitions below the 4502-keV state do not provide
evidence for a measurable lifetime and an upper limit of $\sim$15 ns
can be given for the 4565-keV level.

Angular correlations were used to determine the multipolarity of
some of the newly identified transitions. Because the yield of the
314- and 694-keV isomeric cascade was sufficient, the relevant
coincidence intensities were grouped into 12 different angles
$\theta$. The measured angular-correlation pattern for this pair
strongly favors a sequence with two stretched quadrupole
transitions, as can be seen from the comparison with the theoretical
prediction of Fig.~\ref{fig:fig06}(a), which agrees with a
9/2$^+$$\rightarrow$5/2$^-$$\rightarrow$1/2$^-$ cascade. In view of
the smaller intensities, the correlation between the 1655- and the
1667-keV lines was grouped into five angles
[Fig.~\ref{fig:fig06}(b)]. It is consistent with a quadrupole-dipole
sequence. To be consistent with the decay pattern of the 2663-keV
level, the 1655-keV $\gamma$ ray is proposed as a quadrupole
transition, leading to a 13/2$^+$ assignment for this state, and
15/2$^{+}$ quantum numbers for the level at 4330 keV.

Due to the lack of statistics, the correlation data for other
transitions were regrouped into the two angles of 33$^\circ$ (from
20$^\circ$ to 42$^\circ$ in Gammasphere) and 77$^\circ$ (69$^\circ$
to 87$^\circ$). Intensity ratios were obtained for the 1655-1250,
1655-172, and 1210-1695 keV pairs of transitions. The ratio of
0.81(9) measured for the 1655-1250 keV cascade points to a dipole
character for the 1250-keV $\gamma$ ray, resulting in a 15/2$^+$
assignment to the 3913-keV level. With this 15/2$^+$ assignment and
the measured 1.8(3) ratio indicating a quadrupole-dipole cascade for
the 1210-1695 keV pair where the dipole transition has a large
$E2/M1$ mixing ratio, a consistent picture emerges with the proposed
11/2$^+$ spin and parity for the 2218-keV state. Note that the
mixed-dipole character for the 1210-keV transition is also
consistent with the expectations of shell-model calculations, as
will be discussed below. A 17/2 spin assignment to the 4502-keV
state was derived from the 0.85(4) correlation ratio measured for
the 1655-172 keV pair. Even though correlation data could not be
extracted for the 1210-1312 keV cascade, the 13/2$^+$ assignment to
the 3530-keV level is supported by the presence of the weak,
2522-keV decay branch towards the 9/2$^+$ isomer. Finally, the
general agreement between these assignments and the results of
shell-model calculations was used to tentatively propose 19/2$^-$
and 21/2$^-$ assignments to the two highest states. The experimental
information on levels in $^{67}$Ni is summarized in
Table~\ref{table:tab01}.

    \begin{table*}
    \caption{\label{table:tab01}List of levels with the spin-parity
    assignments and $\gamma$ rays identified in $^{67}$Ni, including
    intensities and placements.}
    \begin{ruledtabular}
    \begin{tabular}{ccccc}
    E$_i$                        & J$_i^{\pi}$   & J$_f^{\pi}$   & E$_{\gamma}$   &   I$_{\gamma}$  \\
    (keV)                        &               &               & (keV)          &                 \\
    \hline
    0                            & 1/2$^{-}$     &               &                &                 \\
    694.3(2)\footnotemark[1]     & 5/2$^{-}$     & 1/2$^{-}$     & 694.3(2)       &                 \\
    1008.1(3)\footnotemark[1]    & 9/2$^{+}$     & 5/2$^{-}$     & 313.8(2)       &                 \\
    2218.0(4)                    & 11/2$^{+}$    & 9/2$^{+}$     & 1210.0(3)      & 66(9)           \\
    2662.8(4)                    & 13/2$^{+}$    & 11/2$^{+}$    & 444.9(3)       & 13(2)           \\
                                 &               & 9/2$^{+}$     & 1654.7(2)      & 100(8)          \\
    3530.3(4)                    & 13/2$^{+}$    & 11/2$^{+}$    & 1312.3(3)      & 16(3)           \\
                                 &               & 9/2$^{+}$     & 2522(1)        & 2.0(5)          \\
    3913.0(4)                    & 15/2$^{+}$    & 13/2$^{+}$    & 382.7(2)       & 16(3)           \\
                                 &               & 13/2$^{+}$    & 1250.0(3)      & 29(5)           \\
                                 &               & 11/2$^{+}$    & 1695.1(5)      & 7(1)            \\
    4330.1(4)                    & 15/2$^{+}$    & 13/2$^{+}$    & 1667.3(2)      & 50(7)           \\
    4501.9(4)                    & 17/2$^{(-)}$  & 15/2$^+$      & 171.8(2)       & 48(6)           \\
                                 &               & 15/2$^+$      & 588.8(2)       & 52(6)           \\
    4564.7(5)                    & (19/2$^-$)    & 17/2$^{(-)}$  & 62.8(2)        & 31(6)           \\
    5273.1(7)                    & (21/2$^-$)    & (19/2$^-$)    & 708.4(5)       & 20(3)           \\
                                 & (21/2$^-$)    & 17/2$^{(-)}$  & 771(1)         & $<$3            \\
    \end{tabular}
    \footnotetext[1]{Observed only with beam off}
    \end{ruledtabular}
    \end{table*}

\section{DISCUSSION}
At first glance, the level structure on top of the 9/2$^+$ isomer in
$^{67}$Ni appears to be of single-particle character. The yrast
sequence does not exhibit any regularity in the increase in
excitation energy with angular momentum, as would be expected in the
presence of collectivity, and states of opposite parity compete for
yrast status. Moreover, in the absence of any notable Doppler shift
for any of the observed transitions, the combined feeding and level
lifetimes must be at least of the order of the stopping time of the
reaction products in the thick uranium target; {\it i.e.} 1 ps or
longer. It should also be noted that the sequence of levels above
the 9/2$^+$ isomer exhibits similarities with the structure found
above the corresponding 9/2$^+$ long-lived state in $^{65}$Ni. The
latter structure was interpreted in terms of single-particle
excitations - see Ref.~\cite{Paw94} for details. These observations
would argue in favor of a subshell closure at $N=40$.

In order to gain further insight into the nature of the observed
$^{67}$Ni states, large-scale calculations were carried out with the
shell-model code ANTOINE \cite{Cau99,Cau04} using both the jj44b
\cite{Lie} and the JUN45 \cite{Hon09} effective interactions. Both
Hamiltonians were restricted to the $f_{5/2}$, $p_{3/2}$, $p_{1/2}$,
and $g_{9/2}$ valence space and assume a $^{56}$Ni core. However,
the required two-body matrix elements and single-particle energies
were obtained from fits to different sets of data. Specifically, the
JUN45 interaction was developed by considering data in nuclei with
$Z\sim 32$ and $N\sim 50$, and excludes explicitly the Ni and Cu
isotopes as the $^{56}$Ni core is viewed as being rather ``soft"
\cite{Hon09}. In contrast, experimental data from $Z=28-30$ isotopes
and $N=48-50$ isotones were incorporated in the fits in the case of
the jj44b interaction \cite{Lie}.

\begin{figure}
        \centering
        \includegraphics[angle=-90,width=0.8\textwidth]{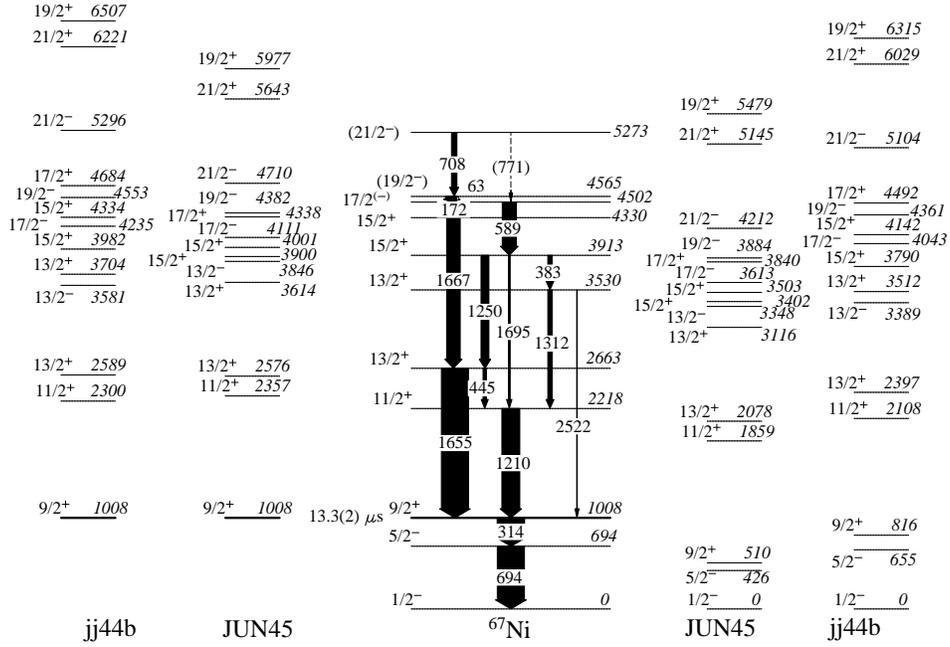}
        \caption{The proposed level scheme of $^{67}$Ni. Results of the
        shell-model calculations with the JUN45 and jj44b effective
        interactions are shown for comparison. The set of
        calculations to the left is identical to that on the right,
        except that the excitation energies were offset such that
        the 9/2$^+$ isomeric state matches the data.}
        \label{fig:fig05}
    \end{figure}

\begin{figure}
        \centering
        \includegraphics[angle=-90,width=0.8\textwidth]{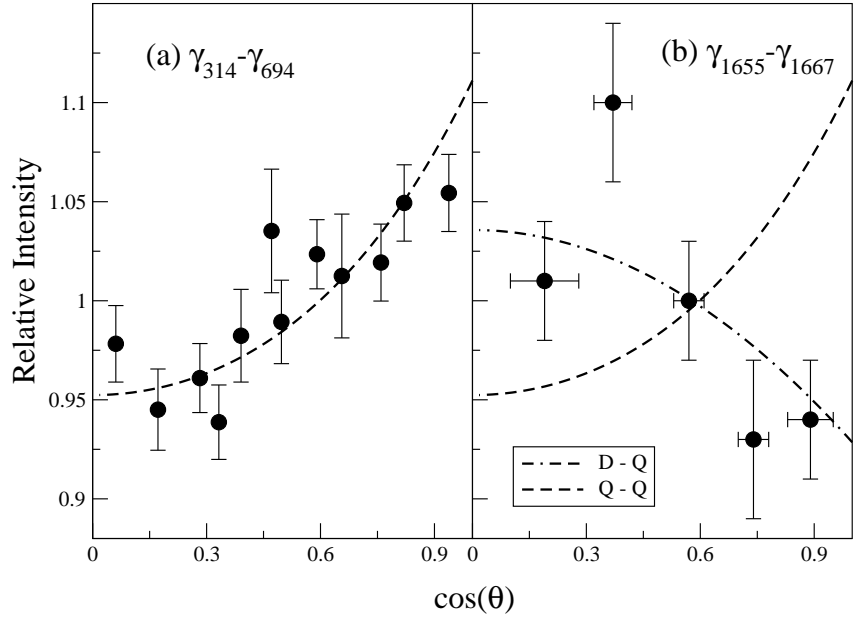}
        \caption{Measured angular correlations for $\gamma\gamma$
        cascades in $^{67}$Ni. Dashed lines in the figure
        correspond to expected patterns associated with pairs of stretched
        quadrupole-quadrupole transitions (panels a and b), while
        the dot-dashed line is associated with a stretched quadrupole-dipole
        pair of $\gamma$ rays (panel b).}
        \label{fig:fig06}
    \end{figure}

The results of the calculations are compared with the experimental
data in Fig.~\ref{fig:fig05}. With both interactions, the energy of
the 9/2$^+$ state is predicted lower than the measured value. This
can be viewed as an indication that the adopted single-particle
energy of the $g_{9/2}$ neutron orbital is too low in the two
Hamiltonians. It is worth noting that the jj44b interaction
calculates this state to lie within 192 keV of the data and
indicates about a 25\% admixture of the $\nu g_{9/2}^3$
configuration into the 9/2$^+$ wave function. With the JUN45
interaction, the level is predicted to lie 498 keV lower than in the
data with roughly 33\% of the wave function involving three neutrons
in the $g_{9/2}$ orbital. This is possibly the result of the
location of the $g_{9/2}$ orbital at a lower energy in the JUN45
Hamiltonian, as compared to that used in the jj44b case, which leads
to larger configuration mixing in the wave function of the $9/2^+$
state.

Overall, the calculated spectrum with both interactions appears
somewhat compressed when compared to the data, as illustrated on the
right side of Fig.~\ref{fig:fig05}. Note that for reasons of
clarity, only the calculated yrast and near-yrast excitations are
shown; {\it i.e.}, the states with a likely corresponding level in
the data are plotted. The correspondence between data and
calculations is rather satisfactory when the computed excitation
energies are expressed relative to the 9/2$^+$ isomer as is done on
the left-hand side of Fig.~\ref{fig:fig05}. Indeed, both
interactions predict close-lying 11/2$^+$ and 13/2$^+$ levels,
separated from the next 13/2$^+$ excitation by roughly 1 MeV, in
agreement with the proposed level scheme. A pair of close-lying
15/2$^+$ levels is also computed to be located directly above the
13/2$_2^+$ state, as seen in the data. Both interactions also
predict a first excited 17/2$^+$ state more than 300 keV above the
15/2$_2^+$ excitation with higher-spin, positive-parity states
another 1.3 MeV or more above this. In contrast, negative-parity
levels are present at lower excitation energies with both effective
interactions, leading to the proposed assignments of 17/2$^{(-)}$,
(19/2$^-$), and (21/2$^-$) for the 4502-, 4565-, and 5273-keV states
in Fig.~\ref{fig:fig05}. As indicated in the figure, these
assignments should be viewed as tentative, especially in the case of
the 17/2 level, where calculated 17/2 states of both parities are
separated only by $\sim$200 and $\sim$400 keV, depending on the
interaction.

It is of interest to identify in the calculations the main
components of the wave functions of the observed states. For the
11/2$^+$ level, and the non-yrast 13/2$_2^+$ and 15/2$_2^+$ states,
both Hamiltonians result in wave functions in which the $\nu
f_{5/2}^5p_{3/2}^4p_{1/2}^1g_{9/2}^1$ configurations dominate with a
contribution of the order of 50\%. Perhaps surprisingly, the
13/2$_1^+$ and 15/2$_1^+$ states are computed to be more fragmented,
with respective main contributions by the $\nu
f_{5/2}^5p_{3/2}^4p_{1/2}^1g_{9/2}^1$ and $\nu
f_{5/2}^4p_{3/2}^4p_{1/2}^2g_{9/2}^1$ configurations of $\sim$30\%
only. In addition, the JUN45 interaction results in a $\sim$10\%
admixture of the $\nu g_{9/2}^3$ configuration into the wave
functions of these two levels. This contribution is of the order of
5\% with the jj44b interaction. With this Hamiltonian the wave
functions of all the negative-parity states are mixed with only the
17/2$^-$ level having a contribution from the  $\nu
f_{5/2}^4p_{3/2}^4p_{1/2}^1g_{9/2}^2$ of the order of 50\%. In
contrast, with the JUN45 interaction, where the ordering of states
is computed in better agreement with the data [see the
15/2$_2^+$\textemdash 17/2$^{(-)}$\textemdash (19/2$^-$)\textemdash
(21/2$^-$) sequence in Fig.~\ref{fig:fig05}], the wave function of
every negative-parity state is characterized by a 40-50\% component
from the $\nu f_{5/2}^4p_{3/2}^4p_{1/2}^1g_{9/2}^2$ configuration.

    \begin{table*}
    \caption{\label{table:tab02}Relative branching ratios
    depopulating the 13/2$_1^+$, 13/2$_2^+$, and 15/2$_1^+$ levels
    derived from experimental measurements, and calculated results
    using the JUN45 and jj44b effective interactions.}
    \begin{ruledtabular}
    \begin{tabular}{ccccccc}
    J$_i^{\pi}$   & J$_f^{\pi}$   & Measurements   & JUN45\footnotemark[1]  & jj44b\footnotemark[1]  & JUN45\footnotemark[2] & jj44b\footnotemark[2]\\
    \hline
    13/2$^+_1$    & 11/2$^{+}$    & 13(2)          & 6      & 23   & 36  & 67  \\
                  & 9/2$^{+}$     & 100(8)         & 100    & 100  & 100 & 100 \\
    \hline
    13/2$^+_2$    & 13/2$^+_1$    & $<$3           & 0      & 166  & 0   & 96  \\
                  & 11/2$^{+}$    & 100(20)        & 100    & 100  & 100 & 100 \\
                  & 9/2$^{+}$     & 11(4)          & 9      & 105  & 7   & 92  \\
    \hline
    15/2$^+_1$    & 13/2$^+_2$    & 57(9)          & 33     & 11   & 95  & 40  \\
                  & 13/2$^+_1$    & 100(15)        & 100    & 100  & 100 & 100 \\
                  & 11/2$^{+}$    & 23(2)          & 15     & 5    & 28  & 7   \\
    \end{tabular}
    \footnotetext[1]{Branching ratios are obtained with calculated transition energies.}
    \footnotetext[2]{Branching ratios are obtained with measured transition energies.}
    \end{ruledtabular}
    \end{table*}

In the absence of lifetime information on the $^{67}$Ni levels above
the isomer, additional tests of the shell-model calculations are
possible by considering the branching ratios for transitions
competing in the deexcitation of specific levels. For the
computation of the $B(E2)$ transition probabilities, proton and
neutron effective charges $e_p=1.5e$ and $e_n=0.5e$ were adopted as
is usual for nuclei in this region. Comparisons between computed
branchings for the two Hamiltonians and the data are presented in
Table~\ref{table:tab02}. Only cases for which the coincidence yields
were sufficient to allow gating on the transitions directly feeding
a state of interest were considered for Table~\ref{table:tab02}.
Note that this table provides shell-model results using either the
calculated or the measured transition energies. The latter values
effectively remove the dependence of the ratios on the transition
energies. From the table, it is clear that calculations with the
JUN45 Hamiltonian are consistently in better agreement with the
measured branching ratios. It is also worth pointing out that both
Hamiltonians also compute a $11/2^+ \rightarrow 9/2^+$ transition of
strongly mixed $E2/M1$ character ($|\delta |> 0.5$), in agreement
with the angular-correlation data for the 1210-1695 keV cascade (see
Section III).

From the discussion above, it is concluded that the levels above the
9/2$^+$ isomeric state can be understood as neutron excitations,
with contributions of protons across the $Z=28$ gap playing a minor
role at best. Calculations with both interactions are in fair
agreement with the data. They attribute a significant role to the
$g_{9/2}$ neutron orbital for every state observed in this
measurement. In fact, in most cases, significant $\nu g_{9/2}^2$ and
$\nu g_{9/2}^3$ configurations are part of the wave functions.
Similar observations have been made for other nuclei close to
$^{68}$Ni; see, for example, recent comparisons between calculations
with the same jj44b and JUN45 interactions and data for $^{65,67}$Cu
in Ref.~\cite{Chi11}. From these findings, it is concluded that even
in a nucleus only one neutron removed from $N=40$, the impact of a
neutron shell closure is rather modest. As the $g_{9/2}$ neutron
orbital is shape driving, multi particle-hole excitations involving
this state may be expected to be associated with enhanced
collectivity and it would be of interest to investigate the latter
in future measurements.

\section{CONCLUSIONS}
A level scheme above the known 13-$\mu$s isomer in $^{67}$Ni was
established for the first time by exploring prompt and delayed
coincidence relationships from deep-inelastic reaction products.
Spin and parity quantum numbers for the newly observed states were
deduced from an angular-correlation analysis whenever sufficient
statistics was available. Shell-model calculations have been carried
out with two modern effective interactions, JUN45 and jj44b, for the
$f_{5/2}pg_{9/2}$ model space with $^{56}$Ni as a core. Satisfactory
agreement between experiment and theory was achieved. Even though
the level structure of $^{67}$Ni appears to exhibit a
single-particle character based on comparisons between the measured
level properties, including branching ratios, with the results of
shell-model calculations, it is suggested that the yrast and
near-yrast states are associated with rather complex configurations.
In fact, calculations indicate that the wave functions of the yrast
states involve a large number of configurations without a dominant
($\sim$50\%) specific one; the latter being more prevalent in the
near-yrast levels. It is hoped that the present data will stimulate
additional theoretical work such as comparisons with calculations
using other effective interactions or a different model space.
Further experimental work aimed at the evolution of the degree of
collectivity with spin and excitation energy is highly desirable as
well.

\begin{acknowledgments}
The authors thank the ATLAS operating staff for the efficient
running of the accelerator and J.P. Greene for target preparation.

This work was supported by the U.S. Department of Energy, Office of
Nuclear Physics, under Contract No. DE-AC02-06CH11357 and Grant No.
DE-FG02-94ER40834, by Polish Scientific Committee Grant No.
2PO3B-074-18, and by Polish Ministry of Science Contract No.
NN202103333.
\end{acknowledgments}

\newpage
\newpage
\newpage
\newpage
\newpage
\newpage
\end{document}